\documentclass[twocolumn,showpacs,floatfix,secnumarabic,amssymb,
amsmath,aps,prl,superscriptaddress,letter]{revtex4}

\usepackage{graphicx}

\begin{document}

\title{Precision Measurement of the Longitudinal Double-spin Asymmetry for Inclusive Jet Production in Polarized Proton Collisions at $\sqrt{s}=200$ GeV}

\affiliation{AGH University of Science and Technology, Cracow, Poland}
\affiliation{Argonne National Laboratory, Argonne, Illinois 60439, USA}
\affiliation{University of Birmingham, Birmingham, United Kingdom}
\affiliation{Brookhaven National Laboratory, Upton, New York 11973, USA}
\affiliation{University of California, Berkeley, California 94720, USA}
\affiliation{University of California, Davis, California 95616, USA}
\affiliation{University of California, Los Angeles, California 90095, USA}
\affiliation{Universidade Estadual de Campinas, Sao Paulo, Brazil}
\affiliation{Central China Normal University (HZNU), Wuhan 430079, China}
\affiliation{University of Illinois at Chicago, Chicago, Illinois 60607, USA}
\affiliation{Cracow University of Technology, Cracow, Poland}
\affiliation{Creighton University, Omaha, Nebraska 68178, USA}
\affiliation{Czech Technical University in Prague, FNSPE, Prague, 115 19, Czech Republic}
\affiliation{Nuclear Physics Institute AS CR, 250 68 \v{R}e\v{z}/Prague, Czech Republic}
\affiliation{Frankfurt Institute for Advanced Studies FIAS, Germany}
\affiliation{Institute of Physics, Bhubaneswar 751005, India}
\affiliation{Indian Institute of Technology, Mumbai, India}
\affiliation{Indiana University, Bloomington, Indiana 47408, USA}
\affiliation{Alikhanov Institute for Theoretical and Experimental Physics, Moscow, Russia}
\affiliation{University of Jammu, Jammu 180001, India}
\affiliation{Joint Institute for Nuclear Research, Dubna, 141 980, Russia}
\affiliation{Kent State University, Kent, Ohio 44242, USA}
\affiliation{University of Kentucky, Lexington, Kentucky, 40506-0055, USA}
\affiliation{Korea Institute of Science and Technology Information, Daejeon, Korea}
\affiliation{Institute of Modern Physics, Lanzhou, China}
\affiliation{Lawrence Berkeley National Laboratory, Berkeley, California 94720, USA}
\affiliation{Massachusetts Institute of Technology, Cambridge, Massachusetts 02139-4307, USA}
\affiliation{Max-Planck-Institut f\"ur Physik, Munich, Germany}
\affiliation{Michigan State University, East Lansing, Michigan 48824, USA}
\affiliation{Moscow Engineering Physics Institute, Moscow Russia}
\affiliation{National Institute of Science Education and Research, Bhubaneswar 751005, India}
\affiliation{Ohio State University, Columbus, Ohio 43210, USA}
\affiliation{Old Dominion University, Norfolk, Virginia 23529, USA}
\affiliation{Institute of Nuclear Physics PAN, Cracow, Poland}
\affiliation{Panjab University, Chandigarh 160014, India}
\affiliation{Pennsylvania State University, University Park, Pennsylvania 16802, USA}
\affiliation{Institute of High Energy Physics, Protvino, Russia}
\affiliation{Purdue University, West Lafayette, Indiana 47907, USA}
\affiliation{Pusan National University, Pusan, Republic of Korea}
\affiliation{University of Rajasthan, Jaipur 302004, India}
\affiliation{Rice University, Houston, Texas 77251, USA}
\affiliation{University of Science and Technology of China, Hefei 230026, China}
\affiliation{Shandong University, Jinan, Shandong 250100, China}
\affiliation{Shanghai Institute of Applied Physics, Shanghai 201800, China}
\affiliation{SUBATECH, Nantes, France}
\affiliation{Temple University, Philadelphia, Pennsylvania 19122, USA}
\affiliation{Texas A\&M University, College Station, Texas 77843, USA}
\affiliation{University of Texas, Austin, Texas 78712, USA}
\affiliation{University of Houston, Houston, Texas 77204, USA}
\affiliation{Tsinghua University, Beijing 100084, China}
\affiliation{United States Naval Academy, Annapolis, Maryland, 21402, USA}
\affiliation{Valparaiso University, Valparaiso, Indiana 46383, USA}
\affiliation{Variable Energy Cyclotron Centre, Kolkata 700064, India}
\affiliation{Warsaw University of Technology, Warsaw, Poland}
\affiliation{University of Washington, Seattle, Washington 98195, USA}
\affiliation{Wayne State University, Detroit, Michigan 48201, USA}
\affiliation{Yale University, New Haven, Connecticut 06520, USA}
\affiliation{University of Zagreb, Zagreb, HR-10002, Croatia}

\author{L.~Adamczyk}\affiliation{AGH University of Science and Technology, Cracow, Poland}
\author{J.~K.~Adkins}\affiliation{University of Kentucky, Lexington, Kentucky, 40506-0055, USA}
\author{G.~Agakishiev}\affiliation{Joint Institute for Nuclear Research, Dubna, 141 980, Russia}
\author{M.~M.~Aggarwal}\affiliation{Panjab University, Chandigarh 160014, India}
\author{Z.~Ahammed}\affiliation{Variable Energy Cyclotron Centre, Kolkata 700064, India}
\author{I.~Alekseev}\affiliation{Alikhanov Institute for Theoretical and Experimental Physics, Moscow, Russia}
\author{J.~Alford}\affiliation{Kent State University, Kent, Ohio 44242, USA}
\author{C.~D.~Anson}\affiliation{Ohio State University, Columbus, Ohio 43210, USA}
\author{A.~Aparin}\affiliation{Joint Institute for Nuclear Research, Dubna, 141 980, Russia}
\author{D.~Arkhipkin}\affiliation{Brookhaven National Laboratory, Upton, New York 11973, USA}
% ...

\collaboration{STAR Collaboration}\noaffiliation

\date{\today}

\begin{abstract}
We report a new high-precision measurement of the mid-rapidity inclusive jet longitudinal double-spin asymmetry, $A_{LL}$, in polarized $pp$ collisions at center-of-mass energy $\sqrt{s}=200$ GeV. The STAR data place stringent constraints on polarized parton distribution functions extracted at next-to-leading order from global analyses of inclusive deep inelastic scattering (DIS), semi-inclusive DIS, and RHIC $pp$ data. The measured asymmetries provide evidence for positive gluon polarization in the Bjorken-$x$ region $x>0.05$.
\end{abstract}


\maketitle


A fundamental and long-standing puzzle in Quantum Chromodynamics (QCD) concerns how the intrinsic spins and orbital angular momenta of
the quarks, anti-quarks, and gluons sum to give the proton spin of $\hbar/2$ \cite{Aidala:2012mv}.  The flavor-summed quark and anti-quark spin contributions, $\Delta\Sigma$, account for less than a third of the total proton spin \cite{deFlorian:2008mr,deFlorian:2009vb,Blumlein:2010rn,Leader:2010rb,Ball:2013lla}. Due to the limited range in momentum transfer at fixed Bjorken-$x$ accessed by fixed-target experiments, the polarized deep-inelastic scattering (DIS) data used to extract $\Delta\Sigma$ provide only loose constraints on the gluon spin contribution, $\Delta{G}$, via scaling violations. 

The measurement of asymmetries directly sensitive to the gluon helicity distribution was a primary motivation for establishing the spin structure program at the Relativisitic Heavy Ion Collider (RHIC).  Since the commencement of the RHIC spin program, several inclusive jet \cite{Abelev:2006uq,Abelev:2007vt,Adamczyk:2012qj} and pion \cite{Adler:2006bd,Adare:2007dg,Adare:2008aa,Adare:2008qb,Adare:2014} asymmetry measurements have been incorporated into next-to-leading-order (NLO) perturbative QCD (pQCD) fits.  While these data provide some constraints on $\Delta{G}$, ruling out large positive or negative gluon contributions to the proton spin, they lack the statistical power to distinguish a moderate gluon contribution from zero. The inclusive jet asymmetries presented here benefit from nearly a 20-fold increase in the event sample as well as improved jet reconstruction and correction techniques compared to \cite{Adamczyk:2012qj}, and provide much tighter constraints on the gluon polarization.

The cross section for mid-rapidity inclusive jet production in $pp$ collisions at $\sqrt{s} = 200$ GeV is well described by NLO pQCD calculations \cite{Jager:2004jh,Mukherjee:2012uz} over the transverse momentum range $5<p_T<50$ GeV/$c$ \cite{Abelev:2006uq}.  The NLO pQCD calculations indicate that mid-rapidity jet production at RHIC is dominated by quark-gluon ($qg$) and gluon-gluon ($gg$) scattering, which together account for $60-90$\% of the total yield for the jet transverse momenta studied here.  The $qg$ and $gg$ scattering cross sections are very sensitive to the longitudinal helicities of the participating partons, so the inclusive jet longitudinal double-spin asymmetry, $A_{LL}$, provides direct sensitivity to the gluon polarization in the proton.  $A_{LL}$ is defined as:
\begin{equation}
  A_{LL} = \frac{\sigma^{++} - \sigma^{+-}}{\sigma^{++} + \sigma^{+-}},
\end{equation}
where $\sigma^{++}$($\sigma^{+-}$) is the differential cross section when the beam protons have the same (opposite) helicities. 

The data presented here were extracted from an integrated luminosity of 20\,pb$^{-1}$ recorded in the year 2009 with the STAR detector \cite{NIM-RHIC} at RHIC. The polarization was measured independently for each of the two counter-rotating proton beams (hereafter designated blue (B) and yellow (Y)) and for each fill using Coulomb-nuclear interference proton-carbon polarimeters \cite{Jinnouchi:2004up}, calibrated via a polarized atomic hydrogen gas-jet target \cite{Okada:2006dd}.  Averaged over RHIC fills,  the luminosity-weighted polarization values for the two beams were $P_B=0.574$ and $P_Y=0.573$, with a $6.5\%$ relative uncertainty on the product $P_BP_Y$ \cite{CNI:2009}. The helicity patterns of the colliding beam bunches were changed between beam fills to minimize systematic uncertainties in the $A_{LL}$ measurement.  Segmented Beam-Beam Counters (BBC) \cite{BBC:2005}, symmetrically located on either side of the STAR interaction point and covering the pseudo-rapidity range $3.4<|\eta|<5.0$, measured the helicity-dependent relative luminosities and served as local polarimeters.

The STAR subsystems used to measure jets are the Time Projection Chamber (TPC) and the Barrel (BEMC) and Endcap (EEMC) Electromagnetic Calorimeters ~\cite{NIM-RHIC}.  The TPC provides tracking for charged particles in the 0.5\,T solenoidal magnetic field with acceptance of $|\eta|<1.3$ and $2 \pi$ in the azimuthal angle $\phi$.  The BEMC and EEMC cover a fiducial area of $-1.0<\eta<2.0$ and $0<\phi<2\pi$, and provide triggering and detection of photons and electrons. 

Events were recorded if they satisfied the jet patch (JP) trigger condition in the BEMC or EEMC. The JP trigger required a $\Delta\eta\times\Delta\phi=1\times{1}$ patch of towers to exceed a transverse energy threshold of $5.4$ (JP1, prescaled) or $7.3$ (JP2) GeV, or two adjacent patches to each exceed $3.5$ GeV (AJP).  The addition of the AJP condition, combined with a reconfiguration of the jet patches so that they overlapped in $\eta$, resulted in a $37\%$ increase in jet acceptance compared to previous data \cite{Adamczyk:2012qj}. Upgrades in the data acquisition system allowed STAR to record events at much higher rates as well.

The analysis procedures were similar to those in \cite{Adamczyk:2012qj} except where noted below.  The inputs to the jet finder were the charged particle momenta measured by the TPC and the neutral energy depositions observed by the calorimeter towers.  Jets were reconstructed using the anti-$k_T$ algorithm \cite{Cacciari:2008gp}, as implemented in the FastJet package \cite{Cacciari:2011ma}, with a resolution parameter $R$ = 0.6.  This is a change from the mid-point cone algorithm \cite{Blazey:2000qt} that was used in previous STAR inclusive jet analyses \cite{Abelev:2006uq,Abelev:2007vt,Adamczyk:2012qj}.  Anti-$k_T$ jets are less susceptible to diffuse soft backgrounds from underlying event and pile-up contributions, which provides a significant reduction in the trigger bias described below.

\begin{figure}[b]
\includegraphics*[width=\linewidth]{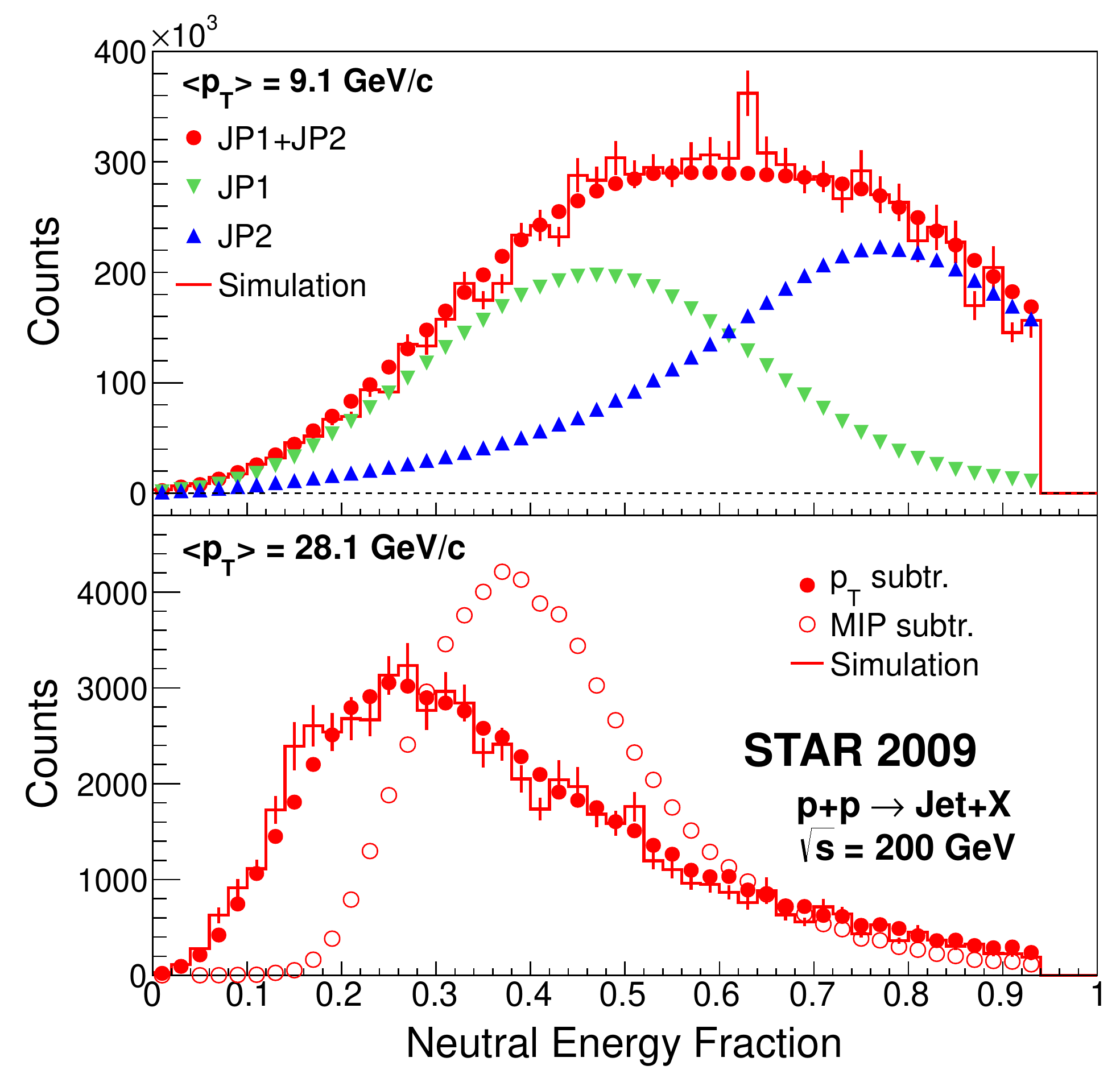}
\caption{\label{fig:jetRt} (Color online.) Jet neutral energy fraction (NEF) comparing data (solid points) with simulations (histograms), where both are calculated with $p_T$ subtraction.  Upper panel shows jets with $8.4<p_T<9.9$ GeV/$c$, demonstrating the bias in NEF when jet $p_T$ is near the trigger threshold.  Lower panel shows jets with $26.8<p_T<31.6$ GeV/$c$, demonstrating an apparent bias persists well above threshold when using MIP subtraction (open circles). The error bars show the simulation statistics. Those for the data are smaller than the points.}
\end{figure}

Most frequently, charged hadrons deposit energy equivalent to a minimum ionizing particle (MIP) in the calorimeter towers. Because the TPC reconstructs the momentum of all charged particles, the inclusion of tower energy from charged hadrons results in an overestimation of the jet momentum.  Fluctuations in the deposited tower energy when charged hadrons interact with calorimeter materials further distort the jet momentum and degrade the jet momentum resolution.  In previous STAR jet analyses \cite{Abelev:2006uq,Abelev:2007vt,Adamczyk:2012qj}, this hadronic energy was accounted for by subtracting energy corresponding to a MIP from the energy deposited in any BEMC or EEMC tower with a charged track passing through it, and then using simulations to estimate the residual correction. In this analysis, the $E_T$ of the matched tower was adjusted by subtracting either $p_{T_{\,}}c$ of the charged track or $E_T$, whichever was less. This procedure reduces the residual jet momentum corrections.  It also reduces the sensitivity to fluctuations in the hadronic energy deposition, resulting in an improved jet momentum resolution of $\simeq$18\% compared to $\simeq$23\% in previous analyses. The bottom panel of Fig.\@ \ref{fig:jetRt} demonstrates that this new ``$p_T$ subtraction" scheme leads to an average for the neutral energy fraction (NEF) of the jet energy that is close to the value of about 1/3 expected from isospin considerations. 

In this analysis, jets were required to have transverse momentum $p_T>5$ GeV/$c$ and $|\eta|<1.0$. Non-collision backgrounds such as beam-gas interactions and cosmic rays, observed as neutral energy deposits in the BEMC and EEMC, were minimized by requiring the NEF to be less than 0.94. Only jets that pointed to a triggered jet patch were considered.  The top panel in Fig.\@ \ref{fig:jetRt} demonstrates the effect of the calorimeter trigger on the jet NEF. The trigger requirement skews the sample to larger neutral energies, especially for jets reconstructed near the trigger threshold. The lower panel shows that this bias is minimized by the $p_T$ subtraction when the jet $p_T$ is well above threshold.

Simulated events are used to calculate the jet momentum corrections and to estimate the systematic uncertainties. This analysis utilized simulated QCD events generated using the Perugia 0 tune \cite{Perugia_ref} in PYTHIA 6.425 \cite{Sjostrand:2006za}.  The PYTHIA events were processed through the STAR detector response package based on GEANT 3 \cite{geant}, and then embedded into randomly triggered events. As a result, the TPC tracks and calorimeter hits reconstructed in the simulation sample incorporate the same beam background and pile-up contributions as the data sample, providing excellent agreement between the data and simulation as shown in Fig. \ref{fig:jetRt}. 

The jet $p_T$ reconstructed at the detector level can be corrected to either the particle or parton level.  Detector jets, which are formed from charged tracks and calorimeter towers, provide contact between the data and simulation. Particle jets are formed from the stable final-state particles produced in a collision.  Parton jets are formed from the hard-scattered partons produced in the collision, including those from initial- and final-state radiation, but not those from the underlying event or beam remnants.  Previous STAR analyses \cite{Abelev:2006uq,Abelev:2007vt,Adamczyk:2012qj} corrected the data back to the particle level.  Here, we correct the data to the parton jet level because parton jets provide a better representation of the jets in an NLO pQCD calculation. The anti-$k_T$ algorithm with $R$ = 0.6 was used to reconstruct parton jets for the simulated PYTHIA events described above. Simulated detector jets were matched to the parton jet closest in $\eta-\phi$ space and within $\sqrt{\Delta\eta^2+\Delta\phi^2}\leq0.5$.  Association probabilities ranged from $76\%$ at the lowest jet $p_T$ to $>98\%$ for $p_T > 9.9$ GeV/$c$.  Asymmetry values are given at the average parton jet $p_T$ for each detector jet $p_T$ bin.

The asymmetry $A_{LL}$ was evaluated according to
\begin{equation}
A_{LL}=
  \frac{
       \sum \left(P_B P_Y\right) \left( N^{++} - r N^{+-} \right)
  }{
       \sum \left(P_B P_Y\right)^2 \left( N^{++}  + r N^{+-} \right)
  },
\end{equation}
in which $P_{B,Y}$ are the measured beam polarizations, $N^{++}$ and $N^{+-}$ denote the inclusive jet yields for equal and opposite proton beam helicity configurations, and $r$ is the relative luminosity.  Each sum is over individual runs that were 10 to 60 minutes long, a period much shorter than typical time variations in critical quantities such as $P_{B,Y}$ and $r$.  Values of $r$ were measured run-by-run, and range from 0.8 to 1.2.  

The STAR trigger biases the data sample by altering the subprocess fractional contributions ($gg$ vs.\@ $qg$ vs.\@ $qq$).  At low $p_T$, the JP efficiency for quark jets is approximately 25\% larger than for gluon jets.  For $p_T>20$ GeV/$c$, the differences are negligible.  Similarly, detector and trigger resolutions may smear and distort the measured $A_{LL}$ values. The size of these effects depends on the value and shape of the polarized gluon distribution as a function of Bjorken-$x$. The $A_{LL}$ values for detector jets were corrected for trigger and reconstruction bias effects by using the simulation to compare the observed asymmetries at the detector and parton jet levels. PYTHIA is not a polarized generator, but asymmetries can be constructed by using the kinematics of the hard interaction to access polarized and unpolarized parton distribution functions (PDFs) and calculate the expected asymmetry on an event-by-event basis.  In this way, the trigger and reconstruction biases were calculated for a range of polarized PDFs that bracket the measured $A_{LL}$ values.  The average of the minimum and maximum $A^{parton}_{LL}-A^{detector}_{LL}$ values for each jet $p_T$ bin was used to correct the measured $A_{LL}$ by amounts ranging from 0.0002 at low $p_T$ to 0.0011 at high $p_T$, and half the difference was assigned as a (correlated) systematic uncertainty. 

\begin{figure}[b]
\includegraphics*[width=\linewidth]{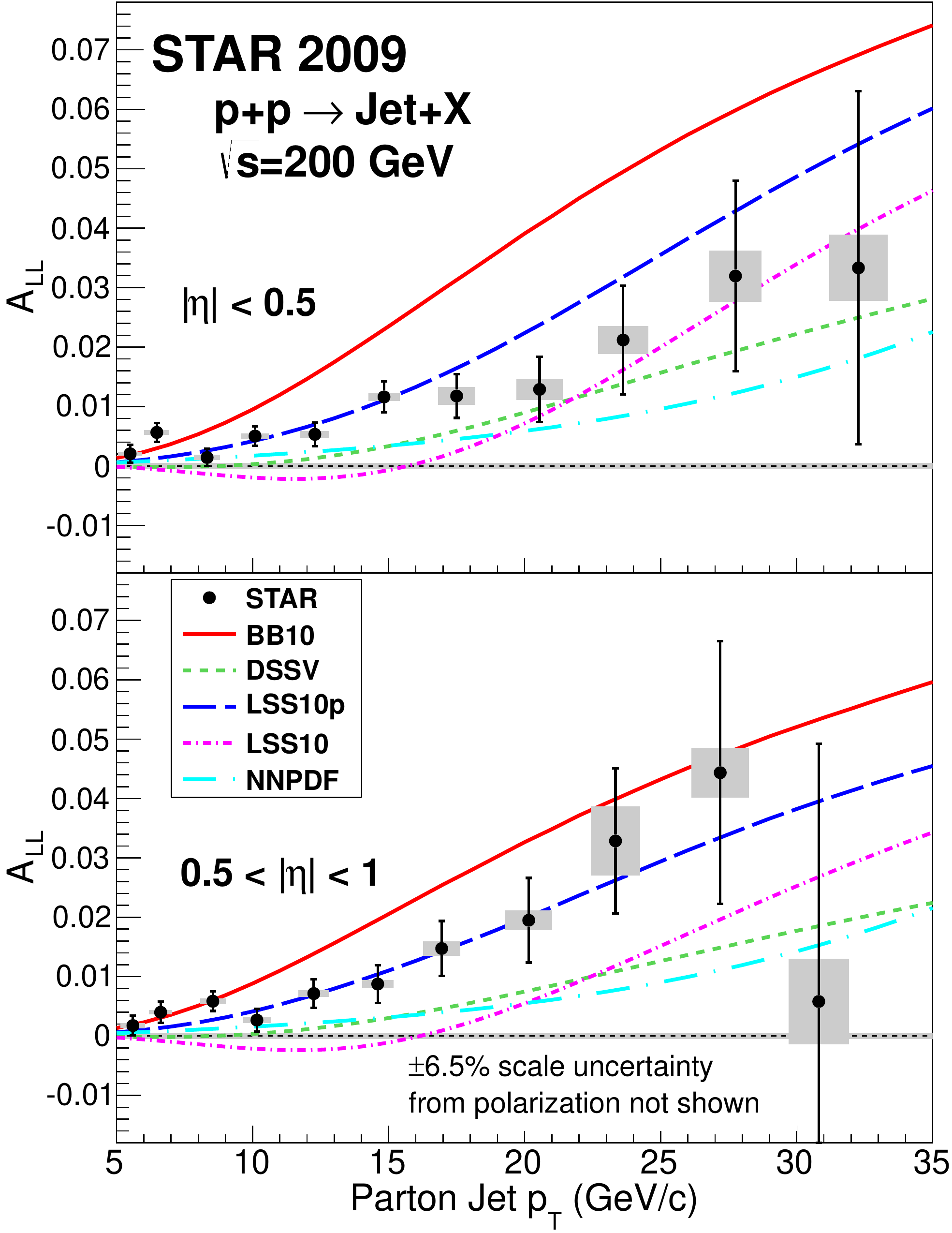}
\caption{\label{fig:jetALL} (Color online.) Midrapidity ($|\eta|<0.5$, upper panel) and forward rapidity ($0.5<|\eta|<1$, lower panel) inclusive jet $A_{LL}$ vs.\@ parton jet $p_T$, compared to predictions from several NLO global analyses.  The error bars are statistical.  The gray boxes show the size of the systematic uncertainties.}
\end{figure}

Figure \ref{fig:jetALL} shows the inclusive jet $A_{LL}$ plotted as a function of parton jet $p_T$ for two $\eta$ bins. The vertical size of the shaded uncertainty bands on the $A_{LL}$ points in Fig. \ref{fig:jetALL} reflects the quadrature sum of the systematic uncertainties due to corrections for the trigger and reconstruction bias ($2-55\times{10}^{-4}$) and asymmetries associated with the residual transverse polarizations of the beams ($3-26\times{10}^{-4}$).  The trigger and reconstruction bias contributions are dominated by the statistics of the simulation sample.  The residual transverse polarization contributions are dominated by the statistical uncertainties in the measurement of the relevant transverse double-spin asymmetry ($A_{\Sigma}$) \cite{Adamczyk:2012qj}.  Both of these uncertainties are primarily point-to-point.  Contributions to $A_{LL}$ from non-collision backgrounds were estimated to be less than 2$\%$ of the statistical uncertainty on $A_{LL}$ for all jet $p_T$ bins and deemed negligible.  The relative luminosity uncertainty ($\pm5\times{10}^{-4}$), which is common to all the points, is shown by the gray bands on the horizontal axes.  It was estimated by comparing the relative luminosities calculated with the BBCs and Zero-Degree Calorimeters \cite{NIM-RHIC}, and from inspection of a number of asymmetries expected to yield null results. The horizontal size of the shaded error bands reflects the systematic uncertainty on the corrected jet $p_T$.  This includes calorimeter tower gain and efficiency and TPC tracking efficiency and momentum resolution effects.  An additional uncertainty has been added in quadrature to account for the difference between the PYTHIA parton jet and NLO pQCD jet cross sections. The PYTHIA vs.\@ NLO pQCD difference dominates for most bins, making the parton jet $p_T$ uncertainties highly correlated

Longitudinal single-spin asymmetries, $A_L$, measure parity-violating effects arising from weak interactions, and hence are expected to be very small compared to $A_{LL}$. $A_L$ was measured and found to be consistent with zero for each beam, as expected for the present data statistics.  

\begin{figure}[b]
\includegraphics*[width=\linewidth]{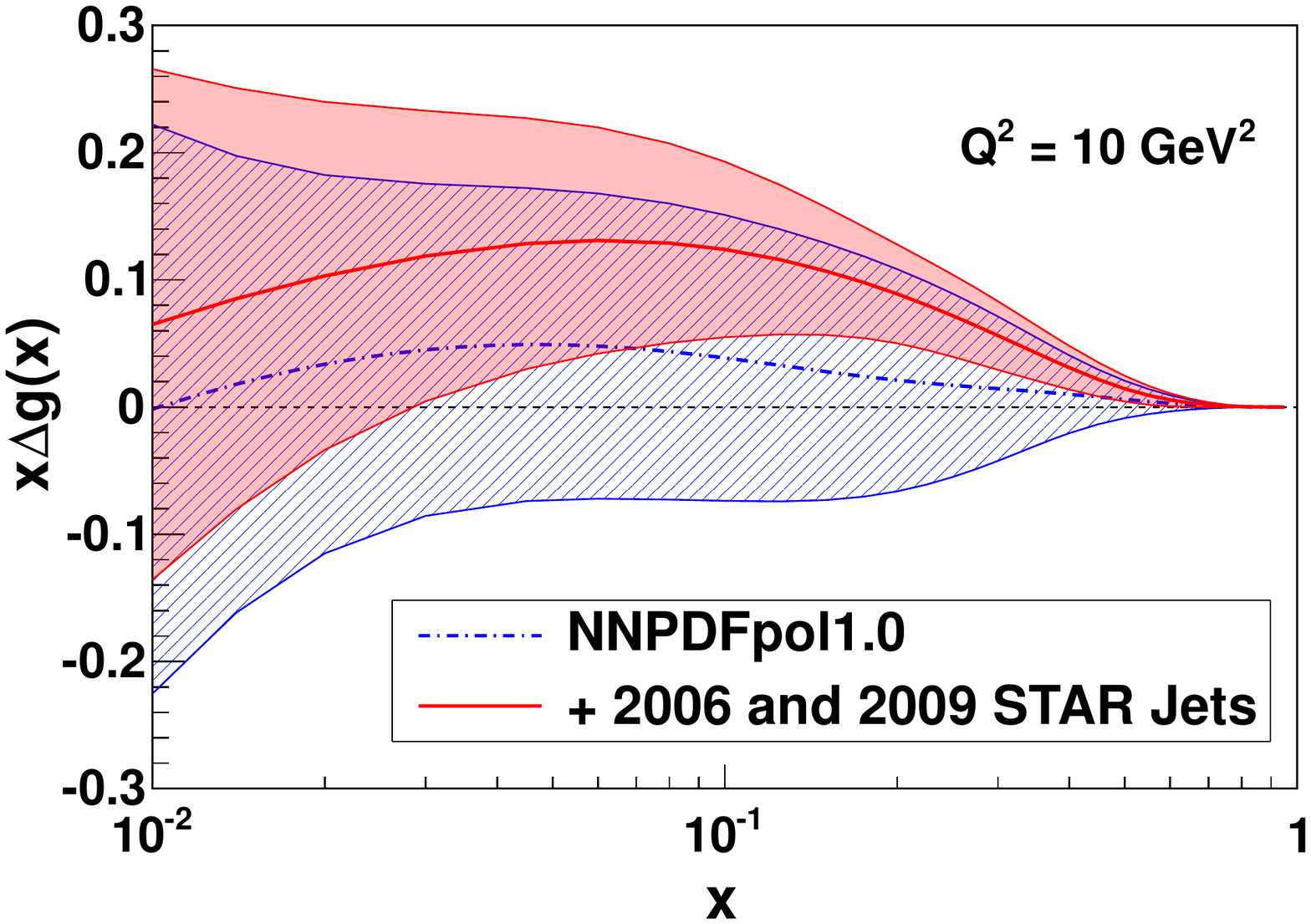}
\caption{\label{fig:NNPDF} (Color online.) Gluon polarizations from NNPDF (blue dot-dashed curve, hatched uncertainty band) \cite{Ball:2013lla}, and from a modified version of NNPDF that we obtain when including the 2006 and 2009 STAR inclusive jet $A_{LL}$ results through reweighting (red solid curve and uncertainty band).}
\end{figure}

The theoretical curves in Fig.\@ \ref{fig:jetALL} illustrate the $A_{LL}$ expected for the polarized PDFs associated with the corresponding global analyses. These predictions were made by inserting the polarized PDFs from BB \cite{Blumlein:2010rn}, DSSV \cite{deFlorian:2008mr,deFlorian:2009vb}, LSS \cite{Leader:2010rb} and NNPDF \cite{Ball:2013lla} into the NLO jet production code of Mukherjee and Vogelsang \cite{Mukherjee:2012uz}.  Theoretical uncertainty bands for $A_{LL}$ were also calculated, but are omitted from the figure for clarity. The BB10 and NNPDF polarized PDFs are based only on inclusive DIS data, while LSS includes both inclusive and semi-inclusive DIS (SIDIS) data sets. LSS provides two distinct solutions for the polarized gluon density of nearly equal quality. The LSS10 gluon density has a node at $x\simeq0.2$, and the LSS10p gluon is positive definite at the input scale $Q^2_0 = 2.5$ GeV$^2$. DSSV is the only fit that incorporates DIS, SIDIS, and previous RHIC $pp$ data.

LSS10p provides a good description of these STAR jet data.  The STAR results lie above the predictions of DSSV and NNPDF and below the predictions of BB10.  However, the measurements fall within the combined data and model uncertainties for these three cases.  In contrast, the STAR jet asymmetries are systematically above the predictions of LSS10 and fall outside the LSS10 uncertainty band for $p_T<15$ GeV/$c$. 

The NNPDF group has developed a reweighting method \cite{Ball:2010gb,Ball:2011gg} to include new experimental data into an existing PDF set without the need to repeat the entire fitting process.  The method involves calculating weighted averages over previously equivalent PDF sets, with the weight for each set derived from the $\chi^2$ probability for the set to describe the new data. We have implemented this method to produce a modified NNPDF fit that includes the 2006 \cite{Adamczyk:2012qj} and 2009 STAR jet data.  When calculating the $\chi^2$ probabilities for the jet asymmetries, we included both the statistical and systematic uncertainties and their correlations.  We find that the jet data have a negligible impact on the polarized quark and anti-quark distributions, but a significant impact on the polarized gluon distribution.  Figure \ref{fig:NNPDF} shows the original NNPDF polarized gluon distribution as a function of $x$ at $Q^2$ = 10 GeV$^2$, as well as the modified fit that includes the 2006 and 2009 STAR data.  The integral of $\Delta{g(x,Q^2=10\,\rm{GeV}^2)}$ over the range $0.05<x<0.5$ is $0.06\pm0.18$ for the original NNPDF fit and $0.21\pm0.10$ when the fit is reweighted using the STAR jet data. The inclusion of the STAR jet data results in a substantial reduction in the uncertainty for the gluon polarization in the region $x>0.05$ and indicates a preference for the gluon helicity contribution to be positive in the RHIC kinematic range.

The DSSV group has performed a new global analysis \cite{DSSV:2014} including the STAR jet $A_{LL}$ results reported in this Letter.  They find that the integral of $\Delta{g(x,Q^2=10\,\rm{GeV}^2)}$ over the range $x>0.05$ is $0.20^{+0.06}_{-0.07}$ at 90\% C.L., consistent with the value we find by reweighting the NNPDF fit.  DSSV indicates that the STAR jet data lead to the positive gluon polarization in the RHIC kinematic range. The functional form of the polarized parton distribution functions assumed by DSSV is less flexible than that assumed by NNPDF, but DSSV also includes substantially more data in their fit.  Both features may contribute to the smaller uncertainty for DSSV relative to NNPDF.

In summary, we report a new high-precision measurement of the inclusive jet longitudinal double-spin asymmetry $A_{LL}$ in polarized $pp$ collisions at $\sqrt{s}=200$ GeV. The results are consistent with predictions from several recent NLO polarized parton distribution fits.  When included in updated global analyses, they provide evidence for positive gluon polarization in the region $x>0.05$. 

We would like to thank J.~Bl{\"u}mlein, H.~B{\"o}ttcher, E.~Leader, E. Nocera, D.~B.~Stamenov,  M.~Stratmann, and W.~Vogelsang for information regarding their respective polarized PDF sets and their uncertainties. We thank the RHIC Operations Group and RCF at BNL, the NERSC Center at LBNL, the KISTI Center in Korea, and the Open Science Grid consortium for providing resources and support. This work was supported in part by the Offices of NP and HEP within the U.S. DOE Office of Science, the U.S. NSF, CNRS/IN2P3, FAPESP CNPq of Brazil,  the Ministry of Education and Science of the Russian Federation, NNSFC, CAS, MoST and MoE of China, the Korean Research Foundation, GA and MSMT of the Czech Republic, FIAS of Germany, DAE, DST, and CSIR of India, the National Science Centre of Poland, National Research Foundation (NRF-2012004024), the Ministry of Science, Education and Sports of the Republic of Croatia, and RosAtom of Russia.


\end{document}